# Surface and finite size effects impact of the phase diagrams, polar and dielectric properties of (Sr,Bi)Ta$_2$O$_9$ ferroelectric nanoparticles

E.A. Eliseev [1], A.V. Semchenko [2], Y.M. Fomichov [1], V.V. Sidsky [2], V.V. Kolos [3],

Yu.M. Pleskachevsky[4], M.V. Silibin[5*], N.V. Morozovsky[6] and A.N. Morozovska[6†],

[1] Institute for Problems of Materials Science, National Academy of Sciences of Ukraine,

Krjijanovskogo 3, 03142 Kyiv, Ukraine

[2] F. Skorina Gomel State University, Sovetskaya 104, Gomel, 246019, Belarus;

[3] JSC «INTEGRAL», Kazintsa 121A, Minsk, 220108, Belarus;

[4] Metal-Polymer Research Institute n. a. V.A. Belyi,

National Academy of Sciences, Belarus,

[5] National Research University of Electronic Technology "MIET",

Bld. 1, Shokin Square, 124498 Moscow, Russia

[6] Institute of Physics, National Academy of Sciences of Ukraine,

46, pr. Nauky, 03028 Kyiv, Ukraine

## Abstract

In the framework of the thermodynamic approach Landau-Ginzburg-Devonshire (LGD) combined with the equations of electrostatics, we investigated the effect of polarization surface screening on finite size effects of the phase diagrams, polar and dielectric properties of ferroelectric nanoparticles of different shapes. We obtained and analyzed the analytical results for the dependences of the ferroelectric phase transition temperature, critical size, spontaneous polarization and thermodynamic coercive field on the shape and size of nanoparticles. The pronounced size effect of these characteristics on the scaling parameter, the ratio of the particle characteristic size to the length of the surface screening, was revealed. Also our modeling predicts a significant impact of the flexo-chemical effect (that is a joint action of flexoelectric effect and chemical pressure) on the temperature of phase transition, polar and dielectric properties of nanoparticles when their chemical composition deviates from the stoichiometric one.

We showed on the example of the stoichiometric nanosized and fine SrBi$_2$Ta$_2$O$_9$ particles that except the vicinity of the critical size, where the system splitting into domains has an important role, results of analytical calculation of the spontaneous polarization have little difference from the numerical ones. We revealed a strong impact of the flexo-chemical effect on the phase transition temperature, polar and dielectric properties of Sr$_y$Bi$_{2+x}$Ta$_2$O$_9$ nanoparticles when the ratio Sr/Bi deviates from the stoichiometric value of 0.5 from 0.35 to 0.65. From the analysis of experimental data we derived the parameters of the theory, namely coefficients of expansion of the LGD functional, contribution of flexo-chemical effect and the length of the surface screening.

---

* corresponding author1, e-mail: anna.n.morozovska@gmail.com

† corresponding author2, e-mail: sil_m@mail.ru

# I. Introduction

Intriguing polar and dielectric properties of ferroelectric nanoparticles attracts the permanent attention of researchers. Yadlovker and Berger [1, 2, 3] present the unexpected experimental results, which reveal the enhancement of polar properties of cylindrical nanoparticles of Rochelle salt. Frey and Payne [4], Zhao et al [5] and Erdem et al [6] demonstrate the possibility to control the temperature of the ferroelectric phase transition, the magnitude and position of the dielectric constant maximum for $BaTiO_3$ and $PbTiO_3$ nanopowders and nanoceramics. The studies of $KTaO_3$ nanopowders [7], $KNbO_3$ and $KTa_{1-x}Nb_xO_3$ nanograins [8, 9, 10] discover the appearance of new polar phases, the shift of phase transition temperature in comparison with bulk crystals. The list of experimental studies of different polar properties of ferroelectric nanoparticles can be continued, however any further comprehensive experimental-and-theoretical study of ferroelectric nanoparticles seems important.

In particular, the surface and finite size effects impact on the phase diagrams, polar and dielectric properties of layered ferroelectric $Sr_yBi_{2+x}Ta_2O_9$ nanoparticles are poorly studied. However the study seems useful for science and advanced applications, because bismuth layer structured ferroelectics (so-called Aurivillius phases with general chemical formulae $Bi_2A_{m-1}B_mO_{3m+3}$ [11, 12, 13]) such as strontium bismuth tantalate, $Sr_yBi_{2+x}Ta_2O_9$ (SBT), vanadate, $Sr_yBi_{2+x}V_2O_9$ and niobate, $Sr_yBi_{2+x}Nb_2O_9$ (SBN) as well as their solid solutions attract permanent scientific interest. These materials, due to their intriguing electronic, ferroelectric and electrophysical properties have been recognized as a prominent candidate for applications in non-volatile ferroelectric memories (NvFRAM) [14, 15, 16, 17, 18] because of negligible fatigue, low leakage currents, and ability to maintain ferroelectricity in the form of thin films [19, 20, 21]. Phase transitions and size effects in $SrBi_2Ta_2O_9$ nanoparticles have been investigated by *in situ* Raman scattering by Yu et al [22] and by thermal analysis and Raman spectroscopy by Ke et al [23]. Yu et al [22] measured the dependence of the transition temperature on the particle size and extrapolated the critical size of the ferroelectricity disappearance as 2.6 nm. Ke et al [23] observed two anomalies in the temperature dependence of specific heat indicating new ferroelectric intermediate phase in the phase-transition sequence and calculated that the nanoparticles are ferroelectric up to the particle size 4.2 nm.

Specifically, many properties of ferroelectric nanoparticles we are interested here (elastic fields, phase transition temperature and polarization reversal process [24]) are controlled by the gradients of defects concentration, polarization and elastic stresses via the Vegard-type chemical stresses [25], flexoelectric and electrostrictive mechanisms [26]. Defects accumulation under a curved surface produces effective stresses of the inner part of the particle due to the lattice expansion or contraction. The characteristic thickness of the layer enriched by defects is determined by the screening length, and their maximal concentration is limited by steric effect [27, 28].

Since the strong gradients of elastic strains can originate in the inner part of the particle from the intrinsic surface stress (surface tension) [29, 30, 31], bond contraction [32, 33] and Vegard stresses [34] they can control the ferroelectric polarization distribution, transition temperature, related polar and dielectric properties via the electrostrictive and flexoelectric mechanisms. The Vegard stresses and strains originate from the host lattice local distortions induced by the drift and diffusion of guest point defects or small clusters, such as mobile light ions, vacancies, interstitial or substitution atoms, etc [35]. The Vegard stress is proportional to the defect concentration variation (i.e. gradient) and the proportionality coefficients are the components of the Vegard tensor [36] (other name elastic dipole). The flexoelectric effect, that is an electric polarization generated in solids by a strain gradient and vice versa [37, 38, 39], can enhance the polar properties. The induced polarization is linearly proportional to the strain gradient and the proportionality coefficients $f$, which are the components of the flexocoupling tensor, are fundamentally quite small, $f \sim e/a$, where $e$ and $a$ are respectively electronic charge and lattice constant [40]. The flexo-chemical effect is a joint action of flexoelectric effect and Vegard-type chemical pressure [41].

Theory of finite size effects in nanoparticles allows one to establish the physical origin of the polar and dielectric properties anomalies, transition temperature and phase diagrams changes appeared under the nanoparticles sizes decrease. In particular, using the continual phenomenological approach Niepce [29], Huang et al [32, 33], Ma [42], Eliseev et al [26] and Morozovska et al [30, 31, 34] have shown, that the changes of the transition temperatures, the enhancement or weakening of polar properties in spherical and cylindrical nanoparticles are conditioned by the various physical mechanisms, such as correlation effect, depolarization field, flexoelectricity, electrostriction, surface tension and Vegard-type chemical pressure.

A brief analytical consideration of these mechanisms impact on the nanoparticles is given in a table I in Ref. [34]. In particular depolarization field always decreases ferroelectric polarization and transition temperature. For majority of cases the depolarization field strength is defined by the surface screening of polarization, however there are only few models describing the effect in nanoparticles [43]. The absence of the complex experimental-and-theoretical studies of the surface screening impact on finite size effects of the phase diagrams, polar and dielectric properties of ferroelectric nanoparticles of different shapes motivated us to perform the present theoretical study and compare results with experimental data for SBT nanoparticles.

## II. LGD-phenomenological description
### A. Problem statement

A thermodynamic LGD equation for polarizations in a uniaxial ferroelectric nanoparticle has a form:

$$\alpha(T)P + \beta P^3 - g\left(\frac{\partial^2}{\partial x^2} + \frac{\partial^2}{\partial y^2} + \frac{\partial^2}{\partial z^2}\right)P = E, \qquad (1)$$

where the coefficient $\alpha$ evidently depends on temperature $T$, $\alpha = \alpha_T(T - T_c)$, nonlinearity coefficient $\beta$ and gradient coefficient g are regarded positive and temperature independent. Here $\mathbf{E} = -\nabla\varphi$ is electric field (sum of external $\mathbf{E}_0$ and depolarization one $\mathbf{E}_d$) that is determined self-consistently from electrostatic problem for electric potential $\varphi$, $\varepsilon_0 \varepsilon_b \Delta\varphi = -\partial P/\partial z$ supplemented by the condition of potential continuity at the particle surface S, $(\varphi_e - \varphi_i)|_S = 0$; the subscript "$i$" means the physical quantity inside the article, "$e$" – outside the particle. Boundary conditions for the polarization $P$ are natural, but the surface screening produced by e. g. ambient free charges at the particle surface S is taken into account:

$$\left.\frac{\partial P}{\partial \mathbf{n}}\right|_S = 0, \quad \left.\left(D_{in} - D_{en} + \varepsilon_0 \frac{\varphi}{\lambda}\right)\right|_S = 0. \qquad (2)$$

Here $\mathbf{n}$ is the outer normal to the particle surface, electric displacement $\mathbf{D} = \varepsilon_0 \varepsilon_b \mathbf{E} + \mathbf{P}$, $\varepsilon_0$ is a universal dielectric constant, $\varepsilon_b$ is a relative dielectric permittivity of background [44], $\lambda$ is the surface screening length that can be much smaller than lattice constant [43].

### B. Analytical solution for a single-domain case

Surface screening leads to the appearance of depolarization field (proportional to the ferroelectric polarization z-component) and to the screening of external field $E_0$ inside the particle. Analytical expressions for electric field were derived for a particle either in paraelectric or single-domain ferroelectric phases for several shapes of the particle. For a sphere of radius $R$ the field is $E^{sphere} = \frac{3\varepsilon_e E_0 - P/\varepsilon_0}{(\varepsilon_b + 2\varepsilon_e + R/\lambda)}$, for infinite cylinder of radius $R$ with ferroelectric polarization normal to the sidewalls, $E^{cyl} = \frac{2\varepsilon_e E_0 - P/\varepsilon_0}{(\varepsilon_b + \varepsilon_e + R/\lambda)}$, and $E^{pill} = \frac{\varepsilon_e E_0 - P/\varepsilon_0}{(\varepsilon_b + \varepsilon_e + h/\lambda)}$ for ultra-thin pill of thickness $h$ the field. These three expressions can be obtained from the interpolation expression for a ferroelectric ellipsoid:

$$E^{ellipsoid} = \frac{\varepsilon_e E_0 - n_d P/\varepsilon_0}{(\varepsilon_b n_d + \varepsilon_e(1 - n_d) + n_d(a/\lambda))} \qquad (3)$$

Here $\varepsilon_b$ and $\varepsilon_e$ are dielectric permittivity of ferroelectric background [44] and external media respectively, $n_d$ is the depolarization factor, depending only on geometry of particles, $a$ is the semi-axis ($R$ or $L$) of ellipsoid in the direction of spontaneous polarization (see **Figure 1a**). Actually equation (3) reproduces exact equations for spheres, cylinders or thin pills at $n_d$=1/3, 1/2 or 1, taking $a = R$ or $a = h$ respectively.

Solution of the problem (1)-(3) gives analytical expression for the critical size $a_{cr}$ and size dependence of the ferroelectric-paraelectric phase transition temperature $T_{cr}$:

$$a_{cr} = \lambda \left( \frac{1}{\alpha_T (T_c - T) \varepsilon_0} - \varepsilon_b - \varepsilon_e \left( \frac{1 - n_d}{n_d} \right) \right), \quad (4)$$

$$T_{cr}(a) = T_c - \frac{n_d}{\alpha_T \varepsilon_0 (\varepsilon_b n_d + \varepsilon_e (1 - n_d) + n_d (a/\lambda))} \quad (5)$$

Here $n_d = \frac{1-\xi^2}{\xi^3}\left(\log\sqrt{\frac{1+\xi}{1-\xi}} - \xi\right)$ is the depolarization factor, the eccentricity ratio $\xi = \sqrt{1-(R/L)^2}$ [45]. The linear dielectric susceptibility, average spontaneous polarization and thermodynamic coercive field are given by expressions $\chi = 1/(\alpha + 3\beta P_S^2)\varepsilon_0$, $P_S = \sqrt{-\alpha/\beta}$ and $E_c = 2\sqrt{-\alpha^3/27\beta}$ respectively. Note, that thermodynamic LGD formalism appertains to single domain properties of a ferroelectric particles, one should not aim to describe quantitatively the coercive field under the presence of realistic polarization reversal via the domain wall motion stages. However, the spontaneous polarization and dielectric susceptibility can be described well enough.

Note, that an additional size dependence of the Curie temperature shift $\Delta T_c$ in comparison with a bulk value $T_C^b$ is possible due to the excess surface stress [29, 34], flexoelectric strains and chemical pressure mechanism [41]. Namely, $T_c(R) = T_C^b + \Delta T_c(R)$, where the surface tension and flexo-chemical contributions for a spherical particle are [41]:

$$\Delta T_c(R) = -\frac{1}{\alpha_T}\left((4Q_{12} + 2Q_{11})\frac{\mu}{R} - \left(F_{sph}^2 + \frac{(4Q_{12} + 2Q_{11})\eta}{(s_{11} + 2s_{12})}\right)\frac{R_0^2}{R^2}\right) \quad (6a)$$

For a prolate ellipsoid

$$\Delta T_c(R, L) = -\frac{1}{\alpha_T}\left(4Q_{12}\mu\left(\frac{1}{R} + \frac{1}{L}\right) - \left(F_{cyl}^2 + \frac{4Q_{12}\eta}{(s_{11} + s_{12})}\right)\frac{R_0^2}{R^2}\right) \quad (6b)$$

Here $T_C^b$ is the bulk Curie temperature, $\mu \sim (1 - 5)$N/m is the positive surface tension coefficient [46, 47], $Q_{ij}$ are the bulk electrostriction tensor coefficients, $s_{ij}$ are elastic compliances modulus of the material, the characteristic size $R_0$ is the thickness of the layer, where accumulated defects gradient create elementary volume changes. $F_{sph}$ and $F_{cyl}$ are the shape-dependent effective flexoelectric strain coefficients. Dimensionless strain $\eta \equiv W\delta N$ is a "compositional" Vegard strain in the polar direction produced by the chemical pressure that can be of arbitrary sign (compressive or tensile) depending on the non-stoichiometry nature (type of impurity ions or vacancies). The absolute value of a Vegard coefficient $W$ for perovskite-like compounds can be estimated as $|W| \propto 10$ Å$^3$ [48]. Note that the both signs of $\Delta T_c$ are possible.

The dependence of the dimensionless transition temperature $T_{cr}/T_C$, spontaneous polarization and thermodynamic coercive field on the ratio of the particle radius to the screening parameter $\lambda$ for ferroelectric particles of different shapes (spheres and cylinders) and various values of dielectric constant of the external environment $\varepsilon_e$ is shown in **Figure 1**. The transition temperature for the cylinders is less than the one for the spheres, since depolarization factor in the latter case (1/3) is less than in the former case (1/2). Since the increase of $\varepsilon_e$ leads to a decrease of the depolarization field (see eq. (3)) critical particle size also decreases whereas $T_{cr}$ increases. The transition temperature and spontaneous polarization tend to the corresponding bulk values with size increase, because the degree of screening increases with the size of the system. Coercive field is not saturated with size increase, because the external field is partially screened by the charges of the surface layer and "active" electric field inside particles decreases with their size increase (at constant external field far from the particle).

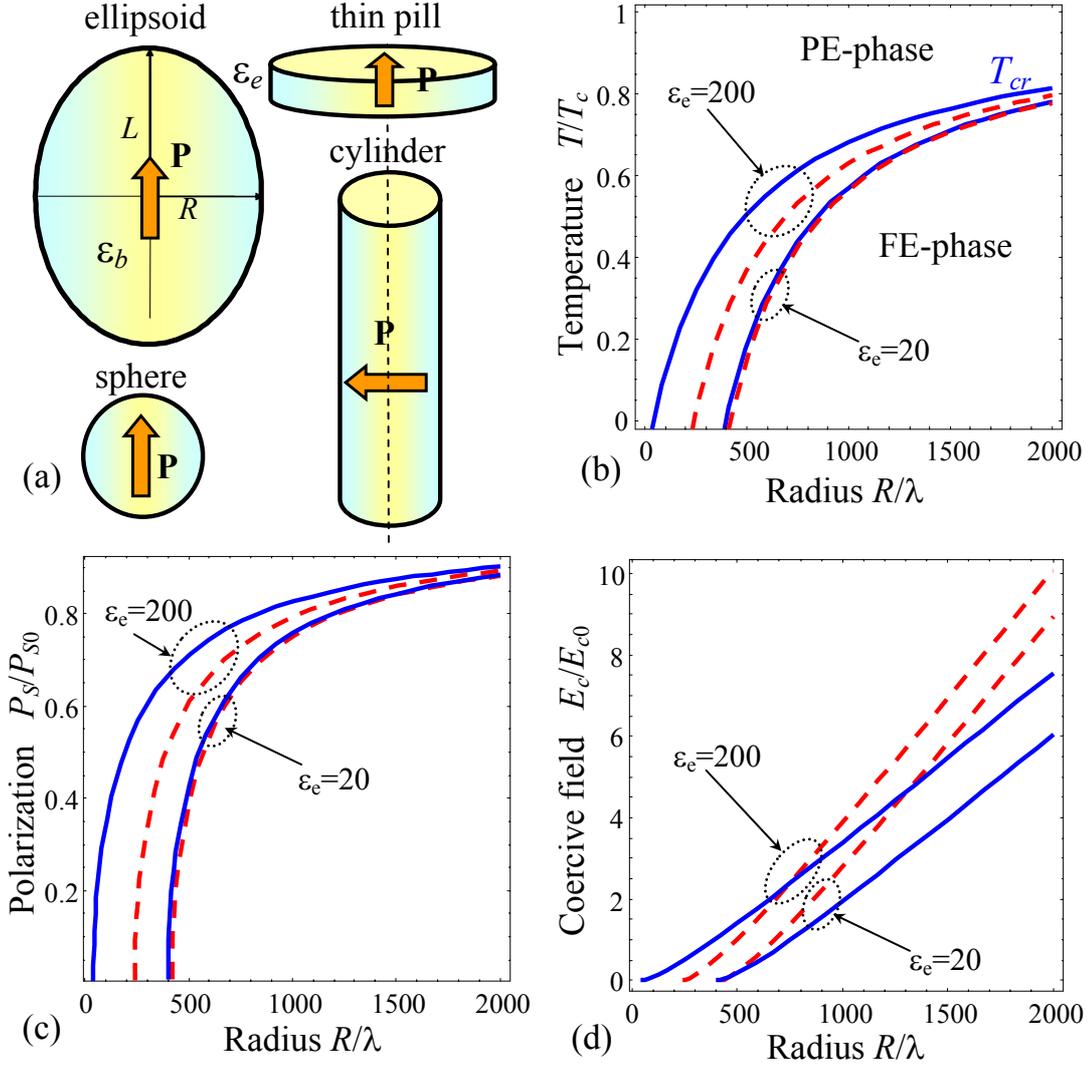

**Figure 1. (a)** Schematics of an ellipsoidal particle with semi-axes $R$ and $L$, sphere, cylinder and thin pill. **(b)** Phase diagram in coordinates temperature $T$ – radius $R$ for the spherical (solid curves) and cylindrical (dashed curves) particles for different values of external media permittivity $\varepsilon_e=20$ and $\varepsilon_e=200$ as marked by ellipses near the curves. Spontaneous polarization of a single-domain particle **(c)** and corresponding thermodynamic coercive field **(d)** as a function of radius for the spherical (solid curves) and cylindrical (dashed curves) particles. Temperature $T=0$, $\varepsilon_e=20$ and $\varepsilon_e=200$ as marked by ellipses near the curves. Normalizing Curie temperature $T_C$, spontaneous polarization $P_{S0}$ and coercive field $E_{c0}$ are corresponding quantities of a bulk material.

### C. Size effects of the phase diagrams and polar properties of SrBi$_2$Ta$_2$O$_9$ nanoparticles

For SrBi$_2$Ta$_2$O$_9$ as bismuth layered-structure ferroelectric, strong anisotropy of physical properties should be inherent due to its structure features consisting in alternate {Bi$_2$O$_2$}$^{2-}$ layers and quasi-perovskite {SrTa$_2$O$_7$}$^{2+}$ blocks [57], the applicability of LGD theory requires additional background. The applicability of LGD approach for layered perovskites SrBi$_2$Ta$_2$O$_9$ [49] and Bi$_5$Ti$_3$FeO$_{15}$ [50] was demonstrated earlier, with the specific remark that a conventional LGD equation (1) can be used for the polarization dynamics in the polar direction normal to the layers, while the materials are almost paraelectric in the planes of the layers. In particular, self-consistent microscopic consideration [51] showed the presence of two very different contributions to ferroelectricity for monocrystalline $SrBi_2Ta_2O_9$, namely the direct contribution through the O-Ta-O bonds in the perovskite blocks $\{SrTa_2O_7\}^{2+}$ and the indirect contributions from the displacement of weakly bounded Bi in $\{Bi_2O_2\}^{2-}$ layers,. These facts allow us to regard $SrBi_2Ta_2O_9$ as a very "rigid" uniaxial ferroelectric and use analytical results of the previous section for its polar properties description.

Effects of the particle shape deviation from the spherical one on the ferroelectric properties of the particles are shown in **Figure 2**, where the transition temperature and spontaneous polarization of $SrBi_2Ta_2O_9$ ellipsoidal particles with different aspect ratio $R/L$ are compared depending on their semi-axis $R$ (in the direction perpendicular to the spontaneous polarization). Here we regard $\Delta T_c(R,L) = 0$ in order to illustrate the net screening effect. It is seen that the extension / contraction of the particles respectively leads to an increase / decrease of the temperature of transition and spontaneous polarization as compared with the properties of the spherical particles (**Figure 2a-c**). The latter is due to the change of the internal depolarization field, which decreases with the particles elongation along the polar axis.

The difference between analytical and numerical calculations of the transition temperature radius dependence (shown in the **Figure 2a**) are absent as anticipated. Note, that the curves' behaviour shown in the **Figure 2a** is in a qualitative agreement with the dependence obtained experimentally by Yu et al [22] (see figure 4 in the ref [22]). Quantitative comparison between the theory and experiment will be presented in the **Figure 3**.

It should be noted that the depolarization field can be significantly reduced with decreasing particle size due to the breaking of the ferroelectric particles into domains with opposite polarization directions, which is manifested as a sharp break on the size dependence of the polarization on the particle radius, calculated by the numerical finite elements method (**Figure 2c**). Except the area near the critical size where the spontaneous polarization appears, where a splitting into domains has an important role (area in the middle of the circle on **Figure 2c**), results of analytical calculation of the spontaneous polarization are close to the numerical ones (compare **Figure 2b** and **2c**). This occurs because with the particle size increase it again becomes monodomain, since the surface screening plays an important role for large particles. This result seems strange and contraries to the facts of the experimental observation of the domain structure in nanoparticles, at first glance. However, it turns out that the large particles transforms into single-domain state only for small λ and the absence of any internal defects (nucleation centers) in our model. The last condition is idealized and almost never attainable in the real physical experiment.

A pronounced widening of the domain walls near the surface of the nanoparticles numerically calculated and shown in **Figure 2d**, is associated with long-range depolarization electric field appearing from the breaks of the double electric layer (consisting of layers of bound and screening charges) at domain wall – interface junctions. The theoretical prediction can be verified by Piezoresponse Force Microscopy (PFM), that is an ideal tool for 3D visualization of the domain structure with nanoscale resolution (see e.g. [52, 53, 54, 55] and refs therein).

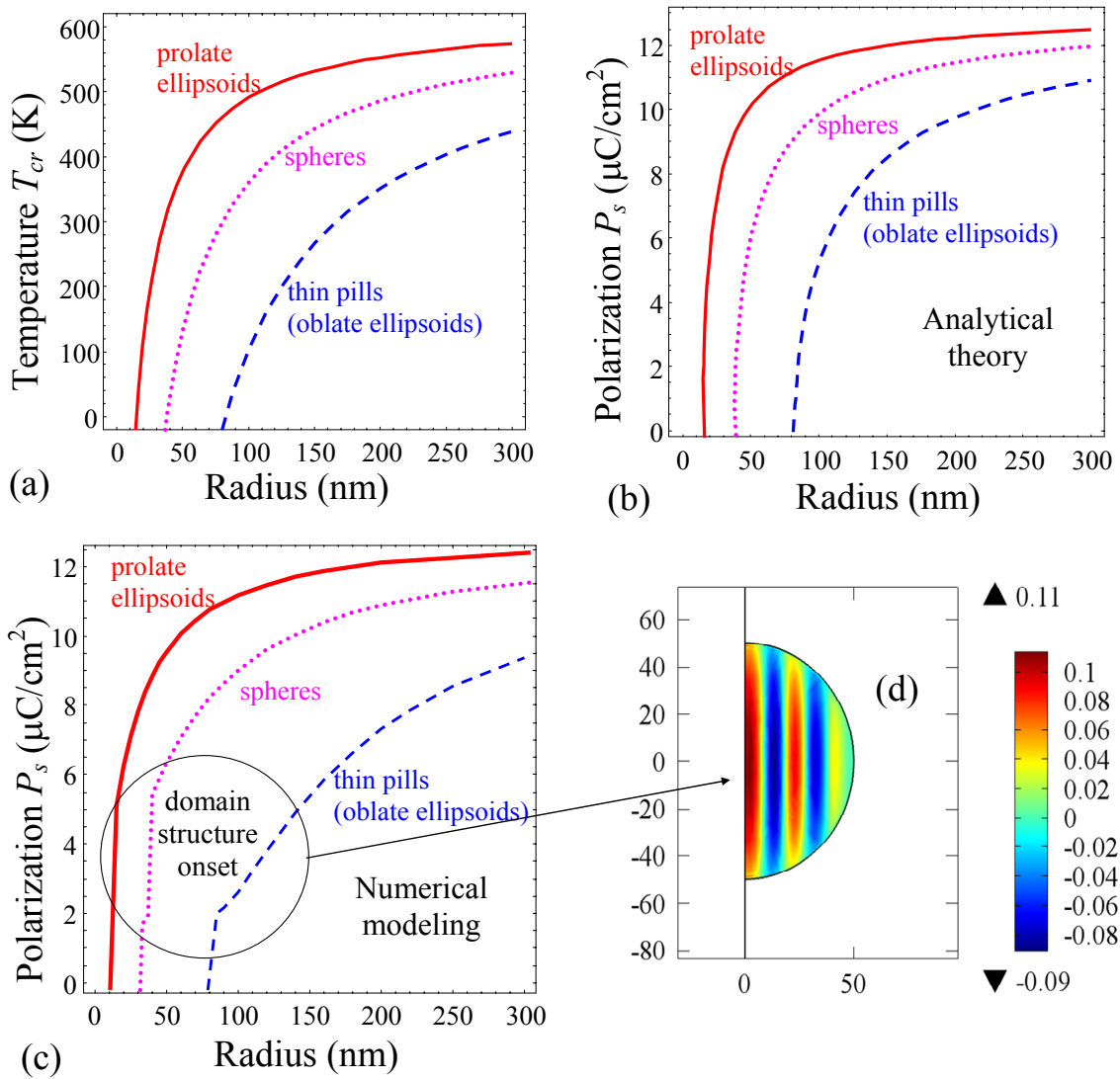

**Figure 2.** Transition temperature **(a)** and spontaneous polarization **(b,c)** of a $SrBi_2Ta_2O_9$ nanoparticle as a function of its radius. **(b)** Analytical theory for a single-domain particle. **(d)** Numerical modeling for particles with possible domain structures, which cross-section is shown in the plot **(d)**. Prolate ellipsoidal shape (solid curves), spherical shape (dotted curve) and thin pills (dashed curves). Material parameters of $SrBi_2Ta_2O_9$ used in the calculations are listed in the **Table I**, most of them were taken from Ref.[49]. The difference between analytical and numerical calculations of the transition temperature are absent as anticipated.

**Table I.** Material parameters of SrBi$_2$Ta$_2$O$_9$ used in the simulations

| coefficient | Numerical value |
|---|---|
| **Symmetry at room $T$** | tetragonal |
| **Background $\varepsilon_b$** | 10 |
| $\alpha$ (C$^{-2}$·mJ) | 4.06($T$–630)×10$^5$ (temperature in Kelvins) |
| $\beta$ (C$^{-4}$·m$^5$J) | 1.5×10$^{10}$ |
| g (m$^3$/F) | 1×10$^{-9}$ |
| $\lambda$ (×10$^{-10}$m) | 1 |

In order to estimate the impact of the surface screening and flexo-chemical effect we fitted the dependence of the experimentally measured [22] ferroelectric transition temperature on the SrBi$_2$Ta$_2$O$_9$ particles' average radius by the following procedure. Using that the approximate equality $(\varepsilon_b n_d + \varepsilon_e(1-n_d) + n_d(a/\lambda)) \approx n_d(a/\lambda)$ is valid with high accuracy, to fit the experimental data we rewrite Eqs.(5)-(6) as following:

$$T_{cr}(R) \approx T_c^b\left(1 - \frac{R_S}{R} - \frac{S_{FC}}{R^2}\right) \qquad (7)$$

Equation (7) contains only 2 fitting parameters, $R_S$ and $S_{FC}$, which signs can be arbitrary, and the dimensions are nm and nm$^2$ correspondingly. The parameter $R_S$ for a sphere is $R_S = \frac{1}{\alpha_T T_c^b}\left((4Q_{12} + 2Q_{11})\mu + \frac{\lambda}{\varepsilon_0}\right)$, for a cylinder it is $R_S = \frac{1}{\alpha_T T_c^b}\left(4Q_{12}\mu + \frac{\lambda}{\varepsilon_0}\right)$. Thus $R_S$ is proportional to the sum of the surface tension coefficient $\mu$ and surface screening length $\lambda$, and so it can be considered as effective surface screening radius. The parameter $S_{FC} = \frac{-1}{\alpha_T T_c^b}\left(F_{sph}^2 + \frac{(4Q_{12} + 2Q_{11})\eta}{(s_{11} + 2s_{12})}\right)R_0^2$ for a sphere and $S_{FC} = \frac{-1}{\alpha_T T_c^b}\left(F_{cyl}^2 + \frac{4Q_{12}\eta}{(s_{11} + s_{12})}\right)R_0^2$ for a cylinder. Parameter $S_{FC}$ is proportional to the flexoelectric constant $F$ and Vegard-type chemical strain $\eta$, and so it can be considered as the effective flexo-chemical radius squire. By fitting the experimental points [22] (shown by solid curves in the **Figure 3**) with Eq.(7) we extracted the parameters $T_c^b$ =588.9 K, $R_S \approx$ 0.315 nm and $S_{FC} \approx$ 4.13 nm$^2$.

Note empirical equation $T_{cr}(R) \approx T_c^b\left(1 - \frac{C}{R - R'}\right)$ proposed for the transition temperature size dependence by Ishikawa et al [56] long ago, did not receive any theoretical background so far. As one can see from the Figure 3 our fitting by Eq.(7) is as good as Ishikawa empirical equation.

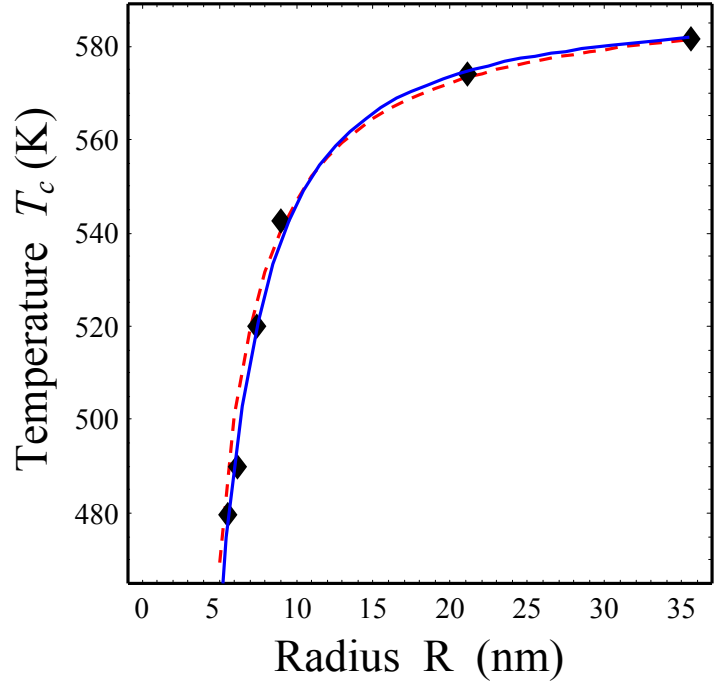

**Figure 3.** Ferroelectric transition temperature on the $SrBi_2Ta_2O_9$ particles' average radius. Diamonds are experimental data from Ref.[22]. Solid curves is the fitting with Eq.(7) for the parameters given in the text, while dashed curve represents empirical Ishikawa equation $T_c = 601(1 - 0.6/(R - 2.1))$ (with radius $R$ in nanometers).

### III. Non-stoichiometry impact on the polar properties of $Sr_{1-x}Bi_{2+x}Ta_2O_9$ particles

The investigated particles and fine-grained ceramics of nonstoichiometric layered strontium bismuth tantalate $Sr_yBi_{2+x}Ta_2O_9$ with different molar ratio Sr:Bi:Ta, were obtained by sol-gel method using the tantalum pentachloride, $TaCl_5$, as one of the starting product together with strontium and bismuth nitrates dissolved in toluene. Nanostructure samples with Sr:Bi:Ta molar ratio 1,4:2,2:2,0; 1,2:2,2:2,0; 1:2,4:2,0; 1,2:2,3:2,0; 0,8:2,2:2,0 were annealed at 750 ° C for 2 hours in oxygen atmosphere.

Ferroelectric hysteresis loops of $Sr_yBi_{2+x}Ta_2O_9$ particles were obtained using a modified precision analyzer of semiconductors parameters HP4156V, probe station Micromanipulator-7000 and control computer with "EasyEXPERT" software. Measurements were carried out under the normal ambient conditions and room temperature. Remanent polarization and coercive field values were defined from the loops. The phase transition temperatures were determined from the maxima of temperature dependences of dielectric permittivity, measured by a digital capacitance measurement circuit. Temperature dependences of dielectric permittivity and P-E hysteresis loops are shown in the **Figure 4**. Dielectric permittivity has a pronounced maximum at the phase transition temperature, and its position shifts to lower temperatures from the sample 1 to the sample 5 (see **Fig. 4a**). Hysteresis loops have a typical ferroelectric shape. The remanent polarization is the highest for the sample 1; a bit lower and almost the same for the samples 2-4, and the smallest for the sample 5 (see **Fig. 4b**). Phase transition temperatures, room temperature dielectric permittivity, remanent polarization and coercive field values are listed in the **Table II**.

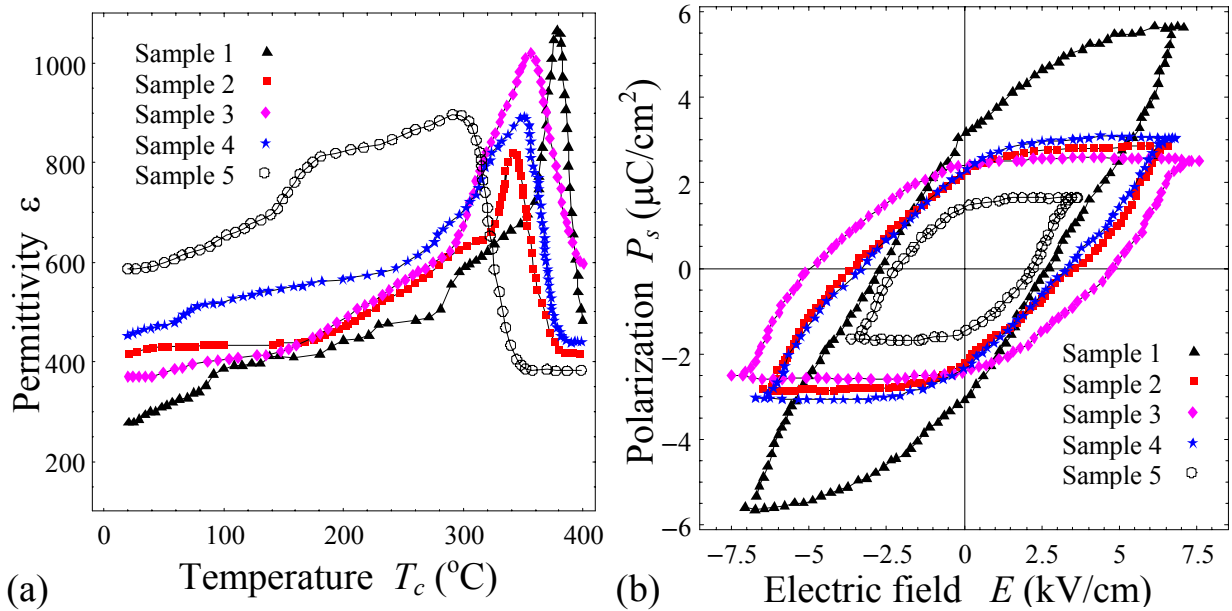

**Figure 4. (a)** Temperature dependences of dielectric permittivity and **(b)** P-E hysteresis loops measured at room temperature (RT) and frequency 1 kHz in SBT nanograined films with a molar ratio Sr:Bi:Ta: 1,4:2,2:2,0 **(sample 1)**; 1,2:2,2:2,0 **(sample 2)**; 1:2,4:2,0 **(sample 3)**; 1,2:2,3:2,0 **(sample 4)**; 0,8:2,2:2,0 **(sample 5)**.

Table II. Parameters of layers tantalate bismuth strontium with different molar ratio Sr:Bi:Ta

| Parameters | SBT samples with different Sr:Bi:Ta molar ratio Sr/Bi=0,64; 0,545; 0,42; 0,52; 0,36 ) | | | | |
|---|---|---|---|---|---|
| | **Sample 1** 1,4:2,2:2,0 Sr/Bi=0.64 | **Sample 2** 1,2:2,2:2,0 Sr/Bi=0.545 | **Sample 3** 1:2,4:2,0 Sr/Bi=0.42 | **Sample 4** 1,2:2,3:2,0 Sr/Bi=0.52 | **Sample 5** 0,8:2,2:2,0 Sr/Bi=0.36 |
| **Particleis' shape** | cylindrical | spherical | spherical | spherical | mixture (sph.+cyl.+aggr.) |
| **Phase transition temperature, $T_{cr}$, ºC measured at 1kHz** | 380 | 340 | 355 | 350 | 295 |
| **Dielectric constant ε at RT, 1kHz** | 277 | 416 | 369 | 453 | 587 |
| **Remanent polarization $P_S$ (µC/cm²) at RT, 1kHz** | 3 | 2.3 | 2.5 | 2.3 | 1.5 |
| **Coercive field $E_c$, kV/cm, at RT, 1kHz** | 2.6 | 3.5 | 4.7 | 3.2 | 2.3 |
| **The average particle size 2R, nm** | 105 | 85 | 109 | 103 | 116 |
| **Roughness, nm** | 5 | 12 | 5 | 11 | 13 |
| **β (×10¹¹ C⁻⁴·m⁵J)** | 2.3 | 2.6 | 2.5 | 2.4 | 4.35 |

| $\alpha_T$ (×10⁵ C⁻²·mJ) | 5.88 | 4.35 | 4.70 | 3.86 | 3.56 |
|---|---|---|---|---|---|
| $\Delta T^*_C$, ºC | 33 | -7 | 8 | 3 | -52 |
| Factor $n_d$ | 1/2 and 0 | 1/3 | 1/3 | 1/3 | from 0 to 1/2 |
| Parameter $R_S$ (nm) | 1 | 2 | 2 | 2 | undefined |
| Parameter $S_{FC}$ (nm²) | 64 | 156 | 156 | 156 | undefined |

As one can see from the **Table II,** the particles shape strongly depends on the molar ratio Sr:Bi:Ta, while the average sizes of $Sr_yBi_{2+x}Ta_2O_9$ particles slightly depend on the ratio. Changes of the $Sr_yBi_{2+x}Ta_2O_9$ particles shape as well as the changes LGD potential expansion coefficients $\alpha_T$ and β under varying the Sr:Bi:Ta molar ratio can be explained by the change of the lattice parameters, its power matrix (bonding forces and rigidity of the lattice) and the surface energy of the material under the composition $x\,y$ change. For example, Idemoto et al [57] performed the corresponding, calculations for the layered system $SrBi_2(Ta_{1-x}Nb_x)_2O_9$ and showed, that doping by Nb or $Bi_2SiO_5$ leads to change of tilt angle of $TaO_6$ octahedra to the *c* axis, octahedron size decrease along the *a* and *b* axes and its increase along the *c* axis, i.e. to anisotropical change of covalent bond parameters. At that the relation between the components of the Vegard's tensor $W_{ij}$, anisotropy of the bulk and the surface free energy coefficients $\alpha_I$ and β should change inevitably. Anisotropy of the surface free energy determines the equilibrium shape of the particles, like the cutting of bulk crystals [58]. Something similar should take place and in the case of $Sr_yBi_{2+x}Ta_2O_9$ particles. On the one hand, self-consistent consideration [59] showed that ferroelectricity of $SrBi_2Ta_2O_9$ originated at least from the perovskite groups $\{SrTa_2O_7\}^{2+}$ and from the weakly bounded Bi in $\{Bi_2O_2\}^{2-}$ layers. On the other hand, it was shown [60] that the formation the "planar" Sr-O defects in the places of discontinuity of Bi-O layers is inherent to $SrBi_2Ta_2O_9$. At that the distance between adjacent perovskite Sr-Ta-O blocks in defective areas is smaller on the 1,2 Å than in the defect-free ones. This implies the presence of the pseudo-perovskite blocks distortions on the borders between defective and defect-free zones in the vicinity of these metal-like defects [60]. It is therefore probable, that due to the strong excess or deficiency of Sr ions for samples 1 and 5 with a ratio Sr/Bi = 0.64 and 0.36 accordingly there is a relative weakening of the bonds in the plane of the layers $\{Bi_2O_2\}^{2-}$ and their gain in the perpendicular direction, containing polar groups $\{SrTa_2O_7\}^{2+}$. This anisotropy (added to natural one [61, 62]) leads to growth of mainly elongated quasi-cylindrical particles in the sample 1 and to the mixture of quasi-spherical, cylindrical and aggregated particles in the sample 5. Small isotropic weakening or strengthening ties due to Bi-Sr interchange occurs seemingly in the samples 2-4 with ratio Sr/Bi close to the stoichiometric ratio 0.5 (see **Table II**). The particles in these samples have preferably quasi-spherical form with a natural facet (see **Figure 3a**). Our hypothesis about the influence of the concentration of the defects on the form of particles is confirmed by the fact that for the samples 1 and 5 containing elongated particles, the effective flexo-chemical parameter $S_{FC}$, that is proportional to the Vegard strain η (see below), appeared in 2.4 times greater than in the samples 2-4 containing quasi spherical particles (see the last row of the table). Furthermore effective parameter $R_S$, that is mainly defined by the surface tension coefficient µ (see below), appeared in 2.6 times greater for the samples 1, contained cylindrical particles (see the penultimate row of the **Table II**).

Dependences of the phase transition temperature, dielectric permittivity, remanent polarization, coercive field and average particle/grain size on the Sr/Bi ratio are shown in the **Figure 5b-e**. Symbols are experimental data for 5 different samples, which polar and dielectric properties are listed in the **Table II**. Solid curves are spline-interpolation functions plotted using the least squares method.

One can see from the **Table II** and from the **Figure 5** that the phase transition temperature, dielectric permittivity and remanent polarization depend on the Sr/Bi ratio in a rather non-trivial and non-monotonic way. At that smaller ε values and greater $P_S$ values correspond to greater $T_{cr}$ values depending of their "remoteness" from $T_{cr}$, which is inherent to ferroelectric state. Other tendency of coercive field changing may be related to pinning phenomenon associated with different state of defects at different Sr/Bi ratios. The average size of the nasnoparticles has minimum at $x \approx 0.55$, indicating the correlation between the size and electrochemical effects. Actually, non-stoichiometry appeared from the deviation of the Sr/Bi ratio $x$ from the stoichiometric value 0.5 can lead to the additional dependence of the phase transition temperature on the ratio $x$ via e.g. excess surface stress and chemical pressure mechanism, e.g. as given by Eqs.(6). The nonlinearity coefficient β also can be $x$-dependent.

An analysis of the experimental results by using analytical expressions (4)-(6) allows us to extract the phenomenological parameters of the model. In particular from equations $\chi(x) + \varepsilon_b = \varepsilon_{\exp}(x)$, $T_{cr} = T_{cr}^{\exp}(x)$ and $P_S = P_r^{\exp}(x)$ at room temperature we extracted the values of λ, $n_d$, β, $\alpha_T$ and $\Delta T_C$, which are listed in the last rows of the **Table II** for all x-ratios of Sr/Bi. At that we regarded that the phase transition temperature of the stoichiometric compound SrBi$_2$Ta$_2$O$_9$ is $T_C^b = 347°C$. Maximal positive deviation from the $T_C^b$ value that is equal to $\Delta T_C = (380 - 347)°C = +33°C$ is observed for the cylindrical particles (sample 1). The positive deviation can originate only from the positive flexo-chemical effect $\left(F_{sph}^2 + \frac{(4Q_{12} + 2Q_{11})\eta}{(s_{11} + 2s_{12})}\right) > 0$, because the positive surface tension can only decrease the transition temperature, since $(4Q_{12} + 2Q_{11})\mu > 0$ for SBT particles of spherical shape.

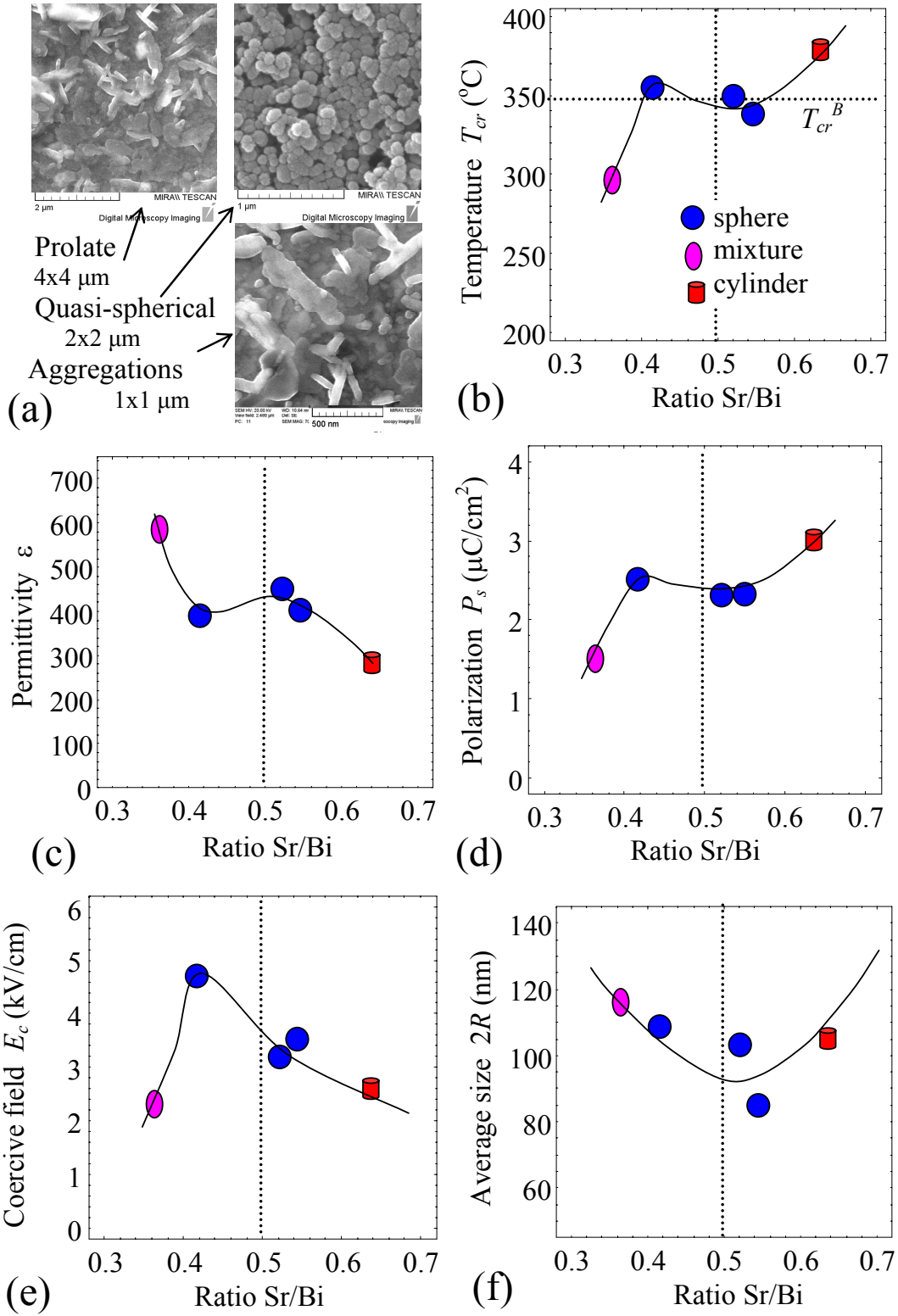

**Figure 5. (a)** STEM images of the studied fine particles of quasi-spherical irregular shape (top, left), and prolate shape (top, right) and their conglomerations (bottom, right). Dependences of the transition temperature **(b)**, dielectric permittivity **(c)**, remanent polarization **(d)**, coercive field **(e)** and average particle/grain size **(f)** on the Sr/Bi ratio. Symbols are experimental data for 5 different samples, which characteristics are listed in the **Table II**. Solid curves are plotted using the least squares method.

In order to estimate the impact of the surface screening and flexo-chemical effect we fitted the dependence of the experimentally measured phase transition temperature on the $Sr_yBi_{2+x}Ta_2O_9$ particles' average radius (shown by symbols in the **Figure 6**) by with Eq.(7) (shown by solid curves in the **Figure 6**). We extracted the parameters $R_S \approx 1$ nm and $S_{FC} \approx 64$ nm$^2$ for spherical and $R_S \approx 2$ nm and $S_{FC} \approx 156$ nm$^2$ for cylindrical particles correspondingly. Since the screening length $\lambda$ is typically very small (<0.1 nm), the difference in the surface screening and flexo-chemical radii for spherical and cylindrical particles originated from the difference in Vegard strain $\eta$ and surface tension coefficients $\mu$ for different Sr/Bi ratios, approving the reasonability of the hypotheses on the defect impact on the particles shape and surface state.

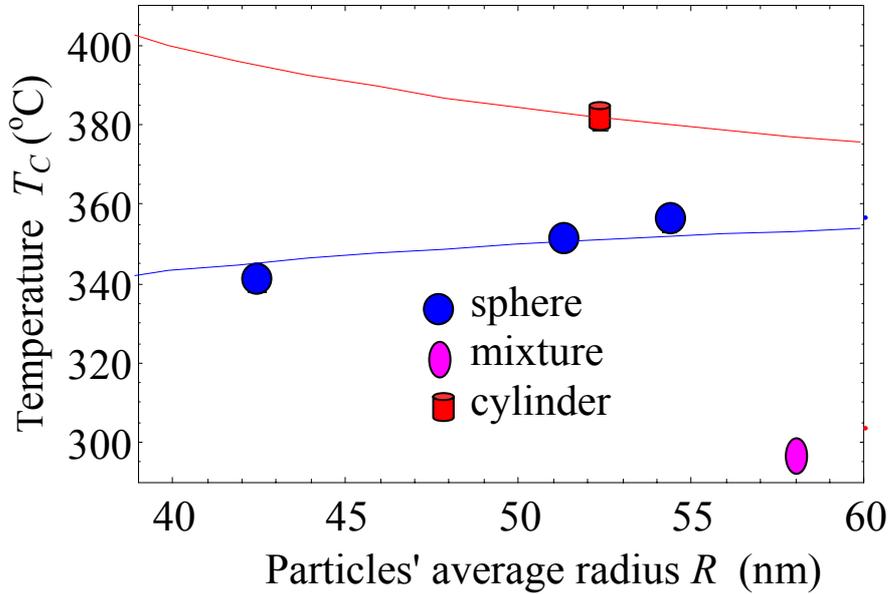

**Figure 6.** Dependence of the phase transition temperature on the particles' average radius. Symbols are experimental data for 5 different samples, which characteristics are listed in the **Table II**. Solid curves are theoretical fitting using the formulae (7) with parameters listed in the last rows in the **Table II**. We could not fit the data for the sample 5, that is in fact admixture of quasi-spherical, elongated quasi-cylindrical and mostly aggregated particles.

### IV. Conclusion

In the framework of the thermodynamic approach Landau-Ginzburg-Devonshire (LGD) combined with the equations of electrostatics, we investigated the effect of polarization surface screening on the size effects of the phase diagrams, polar and dielectric properties of ferroelectric nanoparticles of different shapes. We obtained and analyzed the analytical results for the dependences of the ferroelectric phase transition temperature, critical size, spontaneous polarization and thermodynamic coercive field on the shape and size of nanoparticles. The pronounced dependences of the physical properties on the scaling parameter, the ratio of the particle characteristic size to the length of the surface screening, have been revealed.

We showed on the example of the stoichiometric nanoparticles of $SrBi_2Ta_2O_9$ that except the area near the critical size, where the system splitting into domains has an important role, results of analytical calculation of the spontaneous polarization has little difference from the numerical ones. The proposed model predicts a significant impact of flexoelectric effect and chemical pressure on the temperature of phase transition, polar and dielectric properties of $(Sr,Bi)Ta_2O_9$ particles when a Sr/Bi ratio deviate from the stoichiometric value of 0.5. From the analysis of experimentally obtained data for size effects of $(Sr,Bi)Ta_2O_9$ fine nanoparticles of with Sr/Bi ratio from 0.35 to 0.65 we derived the parameters of the theory (coefficients of expansion of the LGD functional, contribution of flexo-chemical effect and the length of the surface screening).

**Acknowledgements.** E.A.E. and A.N.M. acknowledge National Academy of Sciences of Ukraine (joint Ukraine-Belarus grant 07-06-15) and the Center for Nanophase Materials Sciences, which is a DOE Office of Science User Facility, project CNMS2016-061. A.V.S., V.V.S., V.V.K. and Y.M.P. acknowledge National Academy of Sciences of Belarus grant ФФИ Т15УК/А-67. M.V.S. acknowledges the grant of the President of the Russian Federation for state support of young Russian scientists-PhD (No. 14.Y30.15.2883-MK) and the project part of the State tasks in the field of scientific activity No. 11.2551.2014/K.

**Authors' contribution.** E.A.E. evolved the surface screening model and performed calculations jointly with Y.M.F. A.V.S., V.V.S., V.V.K. and Y.M.P. prepared the samples and obtained experimental results. A.N.M. generated the research idea, acting physical mechanisms choice, and jointly with N.V.M. compared theory with experiment and wrote the manuscript drafts. M.V.S. and N.V.M. jointly with A.N.M. densely worked on the results interpretation, discussion and manuscript improvement.